\documentclass[a4paper,11pt,twoside]{article}

\pdfoutput=1

\makeatletter

\usepackage{amsmath,amssymb}

\usepackage[english]{babel}

\usepackage{cancel}
\usepackage{color}

\usepackage{graphicx}
\usepackage{geometry}

\usepackage{hyperref}

\usepackage[amsmath,hyperref,thmmarks]{ntheorem}

\usepackage{url}

\numberwithin{equation}{section}
\allowdisplaybreaks[1]
\newcommand{\be}{\begin{equation}}
\newcommand{\ee}{\end{equation}}

\newcommand\bbone{{ \mathbb{I}}}

\DeclareMathOperator{\tr}{Tr}

\newcommand{\labitem}[2]{\def\@itemlabel{\textbf{#1}}\item\def\@currentlabel{#1}\label{#2}}

\theoremsymbol{}
\theorembodyfont{\slshape}
\theoremheaderfont{\normalfont\bfseries}
\theoremseparator{}
\newtheorem{Theorem}{Theorem}[section]
\newtheorem{theorem}[Theorem]{Theorem}

\theorembodyfont{\upshape}
\theoremsymbol{\ensuremath{\blacklozenge}}

\theoremstyle{nonumberplain}
\theoremheaderfont{\scshape}
\theorembodyfont{\normalfont}
\theoremsymbol{\ensuremath{\blacksquare}}

\qedsymbol{\ensuremath{_\blacksquare}}
\theoremclass{LaTeX}

\newcommand{\institute}[1]{\newcommand{\@institute}{#1}}

\renewcommand{\maketitle}{
\vspace*{0.5\baselineskip}
{
\center\LARGE\noindent\@title\par
}%
\vspace{1.5\baselineskip}
{
\center\normalsize\noindent\ignorespaces\@author\par
}%
\vspace{0.5\baselineskip}
{
\center\normalsize\ignorespaces\@institute\par
}%
\vspace{2\baselineskip}
}%

\renewenvironment{thebibliography}[1]{%
\section*{References}%
\frenchspacing\small%
\begin{list}{[\arabic{enumi}]}%
{%
\usecounter{enumi}\parsep=2pt\topsep 0pt%
\settowidth{\labelwidth}{[#1]}%
\leftmargin=\labelwidth\advance\leftmargin\labelsep%
\rightmargin=0pt\itemsep=1pt\sloppy%
}%
}{\end{list}}

\begin{document}

\title{Exact Partition Functions for Gauge Theories on $\mathbb{R}^3_\lambda$}

\author{Jean-Christophe Wallet}

\institute{%

\textit{Laboratoire de Physique Th\'eorique (UMR8627)\\
CNRS, University Paris-Sud, University Paris-Saclay, 91405 Orsay, France}\\
e-mail:\href{mailto:jean-christophe.wallet@th.u-psud.fr}{\texttt{jean-christophe.wallet@th.u-psud.fr}}\\[1ex]%
}%

\date{\today}

\maketitle


\begin{abstract} 
\noindent The noncommutative space $\mathbb{R}^3_\lambda$, a deformation of $\mathbb{R}^3$, supports a $3$-parameter family of gauge theory models with gauge-invariant harmonic term, stable vacuum and which are perturbatively finite to all orders. Properties of this family are discussed. The partition function factorizes as an infinite product of reduced partition functions, each one corresponding to the reduced gauge theory on one of the fuzzy spheres entering the decomposition of $\mathbb{R}^3_\lambda$. For a particular sub-family of gauge theories, each reduced partition function is exactly expressible as a ratio of determinants. A relation with integrable 2-D Toda lattice hierarchy is indicated. 
\end{abstract}
\vskip 1 true cm
{\it{To the memory of Raymond Stora}}
\vskip 2 true cm
\noindent Keywords: Noncommutative geometry; differential geometry; gauge theories;\\ renormalisation
\vskip 0,5 true cm
\noindent {\it{2010 Mathematics Subject Classification:}} 81T75; 46L87; 58B34; 81Q80.
\newpage

\section{Introduction}
In Noncommutative Geometry (NCG) \cite{Connes1}, one basic idea is to set-up a kind of duality between spaces and associative algebras so that topological, metric,  differential properties of the space have an algebraic description. A commutative example of such a duality is provided by the Gelfand-Naimark duality between commutative ($C^*$-)algebras and locally compact Hausdorff spaces. When the algebra is no longer commutative, suitable algebraic translation of notions belonging to differential geometry and algebraic topology permits one to define their natural noncommutative analogs so that the noncommutative algebra may be viewed as modeling a ''noncommutative space", in the spirit of the Gelfand-Naimark duality. Many of the building blocks of physics actually fit well with basic concepts of NCG which may ultimately provide efficient tools to improve our understanding of spacetime at short distance. One argument sometimes put forward is that NCG may provide a way to escape physical obstruction to the existence of continuous space-time and commuting coordinates at the Planck scale \cite{Doplich1}. This argument (which however should be only regarded as indicative) has reinforced the interest in noncommutative field theories (NCFT). \par
NCFT appeared in their modern formulation in String field theory \cite{witt1}, followed by models on the fuzzy sphere and almost commutative geometries \cite{mdv1}, \cite{gm90}. NCFT on noncommutative Moyal spaces received a lot of attention from the end of the 90's, in particular from the viewpoint of their renormalisation properties {\footnote{In the following, only NCFT on what could be called informally "totally noncommutative geometries" will be considered, leaving aside the recent developments in gauge models of Connes-Chamseddine-types built on "almost commutative geometries". See e.g \cite{Connes2}, \cite{ccs1}}}. For reviews, see for instance \cite{dnsw-rev}. The renormalisation of NCFT is not an easy task since most of them are non local which precludes the use of any standard treatment devoted to usual local quantum field theories. This may even be complicated by additional peculiarities, among which the so called UV/IR mixing for NCFT built on the popular Moyal space $\mathbb{R}^4_\theta$ which appears already in the $\mathbb{R}$-valued $\varphi^4$ model \cite{minwala}. A family of scalar models, known generically as the Grosse-Wulkenhaar model, was shown to be perturbatively renormalisable to all orders \cite{Grosse:2003aj-pc} at the beginning of the 2000's. Various of its properties have been then investigated \cite{hghs}-\cite{harald-raimar}. Notice that Moyal spaces can support causal structures. Actually, the Moyal plane $\mathbb{R}^2_\theta$ admits a non trivial causal structure between coherent states \cite{causal1}, thus providing a counterexample to some recent claims against any causality in Moyal spaces. Such a causal structure extends very likely to a much wider class of states of $\mathbb{R}^2_\theta$ and to $\mathbb{R}^4_\theta$ as well as to various noncommutative spaces, among which the one considered in this paper which pertains to the group $(C^*)$-algebras. Note that the notion of causality used in \cite{causal1} stems from Lorentzian spectral triple and reduces to the usual notion of causality when the algebra is commutative. NCFT on other noncommutative spaces, such as noncommutative tori, $\kappa$-Minkowski spaces \cite{kappa} have also been considered but their perturbative properties are not so widely explored, in particular for the $\kappa$-Minkowski case due to the present lack of suitable tools able to overcome technical difficulties inherent to the algebraic structure of the $\kappa$-Minkowski algebra. Nevertheless, families of scalar field theories on the noncommutative space $\mathbb{R}^3_\lambda$, a kind of deformation of $\mathbb{R}^3$ introduced a long ago in \cite{Hammaa} (see also \cite{selene}), have been considered recently in \cite{vit-wal-12} and shown to be free of perturbative UV/IR mixing. Among  these, some NCFT were even shown to be finite to all orders in perturbation.\par
At the classical level, the construction of gauge invariant actions is not so difficult, once a differential calculus has been set up, together with a proper notion of noncommutative connection \cite{mdv88-99}, \cite{michor-dbv}.The situation becomes complicated as far as perturbative  behavior is concerned, since gauge invariance requirement supplements the inherent difficulties in the renormalisation of NCFT.  Investigations to extend the features of the Grosse-Wulkenhaar model to a gauge theoretical framework started in the middle of 2000's and produced a lot of articles. This finally gave rise to a gauge invariant model on $\mathbb{R}^4_\theta$ obtained either by effective action computation or by heat kernel methods \cite{Wallet:2007c}. This model appears to be linked to a particular type of spectral triple \cite{finite-vol} whose relationship to the Moyal noncommutative metric geometries \cite{dand-mar-wal} has been analyzed in \cite{Wallet:2011aa}. Unfortunately, its complicated vacuum structure explored in \cite{GWW2} forbids the use of any standard perturbative treatment for $\mathbb{R}^4_\theta$ but not for $\mathbb{R}^2_\theta$, at least for a particular vacuum configuration. This can be achieved by representing the gauge invariant model as a matrix model \cite{MVW13}, showing a relationship with an extension of a 6-vertex model and exhibiting a vacuum instability against quantum fluctuations, as shown in \cite{MVW13}. Alternative approaches based on the implementation of a IR damping mechanism that may render harmless the UV/IR mixing have been also proposed \cite{blaschk1}-\cite{Blaschke:2009c}. This damping approach is appealing. However, interpreting the action within the framework of some noncommutative differential geometry is unclear if possible at all at the present, unlike the case of the induced gauge action. So far,  the construction of a renormalizable gauge theory on $\mathbb{R}^4_\theta$ is unsolved. Another approach provided by the matrix model formulation of noncommutative gauge theories has also evolved partly independently, initiated a long ago in \cite{matrix1} in the context of type IIB (stringy) matrix models. This basically amounts to re-interpret the noncommutative gauge theories as matrix models taking advantage of the relationship between the gauge potential in its noncommutative version and the covariant coordinates (see Section \ref{section2} below). Related works focused on (semi-)classical properties and/or 1-loop computations. For exhaustive reviews on the huge recent literature on this area, see \cite{matrix2} (see also e.g \cite{matrix3}-\cite{matrix5} and references therein).\par
Gauge theories on $\mathbb{R}^3_\lambda$ have been investigated very recently by exploiting the canonical matrix basis introduced in \cite{vit-wal-12} which combined with suitable families of orthogonal polynomials (namely dual Hahn polynomials) and the Favard theorem \cite{toolkit}, a corollary of the spectral theorem, leads to a tractable computation of the relevant propagator. These investigations on gauge theories were partly motivated by the absence of UV/IR mixing and the occurrence of a natural UV cut-off in families of  NCFT studied in \cite{vit-wal-12}, stemming from the very structure of the $\mathbb{R}^3_\lambda$ algebra. In \cite{gervitwal-13}, a family of gauge theories on $\mathbb{R}^3_\lambda$ , which may be viewed as describing the fluctuations of the gauge potential around the classical vacuum $A_\mu=0$, was shown to exhibit the mild perturbative UV behavior expected from \cite{vit-wal-12}. However, the classical vacuum for this family is unstable against quantum fluctuations as shown in \cite{gervitwal-13}. It turns out that some of these gauge theory models, when truncated to a single "fuzzy sphere" $\mathbb{M}_{2j+1}(\mathbb{C})$, can be related to a particular version of the Alekseev-Recknagel-Schomerus action \cite{ARS}, which pertains to the area of string theory and describes a low energy action for brane dynamics on $\mathbb{S}^3$. In \cite{GJW2015}, a family of gauge theories on $\mathbb{R}^3_\lambda$ in a different background corresponding to the so called gauge-invariant connection has been considered and shown to be UV finite to all orders in perturbation and without any IR singularity.\par 
This family of perturbatively finite gauge theories indexed by 3 positive parameters will be the subject of the present paper. In Section \ref{section2}, all the noncommutative data fixing the structure of the classical action are given and discussed, outlining the essential ingredients and possible ways of extensions. A particular emphasis is put on the presentation of the algebra $\mathbb{R}^3_\lambda$ that does not resort on star products and related machinery of deformation theory. The gauge-invariant connection occurring in the specific differential calculus chosen here is discussed. In Section \ref{section3}, the main properties of the family of gauge-fixed actions are outlined and discussed. The partition function factorizes into an infinite product of factors, each of these factors, says $Z_n$, corresponding to the partition function of the gauge theory truncated to $\mathbb{M}_{n}(\mathbb{C})$, $n\in\mathbb{N}$. Fixing one parameter to a specific value modifies the quartic interaction term. Then, each $Z_n$ is shown to be expressible as a ratio of determinants so that the corresponding truncated gauge theory is solvable. A relation to integrable 2-D Toda lattice hierarchy and some reduction is indicated. Section \ref{section4} summarizes the results. 
\section{\texorpdfstring{Gauge theory models on $\mathbb{R}^3_\lambda$}{Gauge theory models.}}\label{section2}
\subsection{\texorpdfstring{$\mathbb{R}^3_\lambda$ and group algebras.}{Group algebras}}\label{section21}
For the ensuing analysis, the algebra $\mathbb{R}^3_\lambda$ can be conveniently presented as
\begin{equation}
\mathbb{R}^3_\lambda = (\bigoplus_{j\in\frac{\mathbb{N}}{2}} \ \mathbb{M}_{2j+1}(\mathbb{C}),.)\label{convolution-alg}
\end{equation}
where $\mathbb{M}_{2j+1}(\mathbb{C})$ is the algebra of $(2j+1)$x$(2j+1)$ complex matrices and the symbol "$.$" denotes the usual operator ("matrix") product which will not be explicitly written in the following. $\mathbb{R}^3_\lambda$ is obviously unital with involution defined by the (hermitean) conjugation. Recall that $\mathbb{M}_{2j+1}(\mathbb{C})$ is referred in the physics literature as the algebra of fuzzy sphere of radius $j$. Hence, $\mathbb{R}^3_\lambda$ can be viewed informally as an infinite sum of fuzzy spheres. A more precise (albeit less intuitive) characterization of this noncommutative space may be obtained from considerations of harmonic analysis on $SU(2)$.\par 

Indeed, from \eqref{convolution-alg}, it can be readily observed that the infinite direct sum decomposition coincides with the Peter-Weyl decomposition of $L^2(SU(2))$ which therefore shares its linear structure with $\mathbb{R}^3_\lambda$. Recall that for any compact (topological) group $G$, one can write  $L^2(G)=\oplus_{\chi\in\hat{G}}E_\chi$ where $\hat{G}$ is the countable set of equivalence classes of irreducible representations of $G$ and $E_\chi$ is the vector space of coefficients of the representation $\chi$, i.e the vector space generated by $\langle \chi(g)u,v\rangle$, $u,v$ being vectors of the representation space of $\chi$. Moreover, the vector space $E_\chi$, endowed with the convolution product on $G$ is an algebra, isomorphic to $\mathbb{M}_n(\mathbb{C})$ with $n=\dim(\chi)$. In the $G=SU(2)$ case, one has the decomposition $L^2(SU(2))=\bigoplus_{j\in\frac{\mathbb{N}}{2}}\mathbb{M}_{2j+1}(\mathbb{C})$. The $SU(2)$ Fourier transform defines a map
\begin{equation}
\mathcal{F}:L^2(SU(2))\to\bigoplus_{j\in\frac{\mathbb{N}}{2}}\mathbb{M}_{2j+1}(\mathbb{C}),\ \ \hat{f}:=\mathcal{F}(f)=\oplus_{j\in\frac{\mathbb{N}}{2}}\int_{SU(2)}d\mu(x)f(x)t^j(x^{-1})\label{fourier}
\end{equation}
for any function $f\in L^2(SU(2))$ where $d\mu(x)$ is the Haar probability measure for $SU(2)$. Here, $t^j(x)$ is the so called matrix of the coefficients of the representation for $x\in SU(2)$ whose elements are given (in obvious notations) by $(t^j(x))_{mn}=\langle jm|\chi_j(x)|jn\rangle$ where $\{|jm\rangle\}_{-j\le m\le j} $ is the orthonormal family spanning the carrier space of the representation indexed by $j\in\frac{\mathbb{N}}{2}$, which is nothing but a Wigner $D$-matrix. The inverse map is
\begin{equation}
\mathcal{F}^{-1}:\bigoplus_{j\in\frac{\mathbb{N}}{2}}\mathbb{M}_{2j+1}(\mathbb{C})\to L^2(SU(2)),\ \ \mathcal{F}^{-1}(\hat{f})(x)=\oplus_{j\in\frac{\mathbb{\mathbb{N}}}{2}}(2j+1)\mbox{tr}_j(t^j(x)\hat{f})\label{fourier-1}
\end{equation}
where $\mbox{tr}_j$ is the canonical trace on $\mathbb{M}_{2j+1}(\mathbb{C})$ for any $j\in\frac{\mathbb{N}}{2}$.\\
In this framework, $\mathbb{R}^3_\lambda$ may be naturally interpreted as the (Fourier transform of the) convolution algebra of $SU(2)$, i.e the Fourier transform as given by relation \eqref{fourier} of $(L^2(SU(2)), \bullet)$ where $\bullet$ is the associative convolution product on $SU(2)$ given for any functions $f,g\in L^1(SU(2))$ by $f\bullet g(u)=\int_{SU(2)} d\mu(t)f(ut^{-1}) g(t)$. Other interesting larger group algebras can be obtained from the convolution algebra, namely the von Neumann algebra $\mathcal{A}(SU(2))$, i.e the multiplier algebra of $C^*(SU(2))$ which is the $C^*$-algebra of the group $SU(2)$. They will not be needed here.\\

In order to make connection with the physics literature, one can notice that this structure singles out natural "coordinates" given by the (hermitean) generators $x_\mu$, $\mu=1,2,3$ of the Lie algebra $su(2)$ which can be expressed conveniently within a suitable basis for $\mathbb{R}^3_\lambda$. For any $j\in\frac{\mathbb{N}}{2}$, let $\{v^j_{mn} \}$, $-j\le m,n\le j$ denotes the canonical basis for $\mathbb{M}_{2j+1}(\mathbb{C})$. Hence, $\mathbb{R}^3_\lambda$ inherits a natural orthogonal basis given by
\begin{equation}
\{v^j_{mn}\},\ -j\le m,n\le j,\ j\in\frac{\mathbb{N}}{2}\label{canonic-base},
\end{equation}
with
\begin{equation}
(v^j_{mn})^\dag=v^j_{nm},\ v^{j_1}_{mn} v^{j_2}_{qp} = \delta^{j_1j_2} \delta_{nq} \ v^{j_1}_{mp},\ -j_1\le m,n\le j_1,\ -j_2\le p,q\le j_2 \ \label{fusion-rules}
\end{equation}
for any $j,j_1,j_2\in\frac{\mathbb{N}}{2}$. Here, orthogonality of the basis \eqref{canonic-base} holds with respect to the hermitean product $\langle a,b \rangle:=\tr(a^\dag b)$ where the trace for any $a,b\in\mathbb{R}^3_\lambda$ is
\begin{equation}
\tr(ab) := 8 \pi \lambda^3 \sum_{j\in\frac{\mathbb{N}}{2}} (2j+1) \mbox{ tr}_j(A^jB^j)\label{trace},
\end{equation}
according to \eqref{fourier-1}, where $A^j\in\mathbb{M}_{2j+1}(\mathbb{C})$ is the matrix arising in the blockwise expansion of $a\in\mathbb{R}^3_\lambda$ in the basis \eqref{canonic-base}
\begin{equation}
a= \sum_{j\in\frac{\mathbb{N}}{2}} \ \sum_{-j\le m,n\le j} a^j_{mn} \ v^j_{mn} \ , \label{nat-fourier}
\end{equation}
so that $(A^j)_{mn}=a^j_{mn}$ (and similarly for $B^j$). The overall factor in \eqref{trace}, where $\lambda$ has mass dimension $[\lambda]=-1$, has been installed for further convenience.\par 

The unit of $\mathbb{R}^3_\lambda$ can be written as
\begin{equation}
\bbone=\sum_{j\in\frac{\mathbb{N}}{2}}P_j,\ P_j=\sum_{m=-j}^jv^j_{mm}\label{unit}
\end{equation}
where for any $j\in\frac{\mathbb{\mathbb{N}}}{2}$, $P_j$ is the orthogonal projector on $\mathbb{M}_{2j+1}(\mathbb{C})$. One easily obtains
\begin{equation}
\mbox{tr}_j(v^j_{mn}) = \delta_{mn} \ , \ \
\langle v^{j_1}_{mn} , v^{j_2}_{pq} \rangle = 8 \pi \lambda^3 \sum_{j_1\in\frac{\mathbb{N}}{2}} w(j_1) \ \delta^{j_1j_2} \delta_{mp} \delta_{nq} \ . \label{orthonormal}
\end{equation}
As a remark, notice that one has $V_S:=8\pi\lambda^3\mbox{tr}_j(P_j)=8\pi\lambda^3(2j+1)^2$ so that summing over $j$ up to, says, $J$ using \eqref{trace} yields $V_S=8\pi\lambda^3\sum_{k=0}^J(k+1)^2\sim\frac{4}{3}\pi(\lambda J)^3$ which mimics the volume of a sphere of radius $\lambda J$. Notice also that the trace \eqref{trace} is almost similar to the trace considered in \cite{gervitwal-13} and \cite{GJW2015} whose choice was partly done from algebraic considerations.\\

The center of $\mathbb{R}^3_\lambda$, $\mathcal{Z}(\mathbb{R}^3_\lambda)$ is the set of the elements of $\mathbb{R}^3_\lambda$ having the following expansion
\begin{equation}
z=\sum_{j\in\frac{\mathbb{N}}{2}}f(j)P_j\label{center},
\end{equation}
where $f(j)$ can be (formally) expanded in $j$ so that $\mathcal{Z}(\mathbb{R}^3_\lambda)$ is actually generated by
\begin{equation}
x_0=\lambda\sum_{j\in\frac{\mathbb{N}}{2}}jP_j\label{x0}
\end{equation}
which is often referred in the physics literature as the radius operator. Note that the overall factor $\lambda$ in \eqref{x0} yields $[x_0]=-1$. \\
From \eqref{convolution-alg} and \eqref{canonic-base}, one infers that
\begin{eqnarray}
x_1&=&\frac{\lambda}{2}\sum_{j,m}\big(\sqrt{(j+m)(j-m+1)}v^j_{m,m-1}+\sqrt{(j-m)(j+m+1)}v^j_{m,m+1}\big),\label{x1}\\
x_2&=&\frac{\lambda}{i2}\sum_{j,m}\big(\sqrt{(j+m)(j-m+1)}v^j_{m,m-1}-\sqrt{(j-m)(j+m+1)}v^j_{m,m+1}\big),\label{x2}
\end{eqnarray}
\begin{equation}
x_3={\lambda}\sum_{j,m}mv^j_{mm},\label{x3}
\end{equation}
from which, by using \eqref{fusion-rules} one obtains
\begin{equation}
[x_\mu,x_\nu]=i\lambda\varepsilon_{\mu\nu\rho}x_\rho,\ [x_\mu,x_0]=0,\ \forall \mu,\nu,\rho=1,2,3\ \label{def-com-rel1}
\end{equation}
\begin{equation}
x_0^2+\lambda x_0=\sum_{\mu=1}^3x_i^2, \label{def-com-rel2}
\end{equation}
which reproduce the "defining relations" of $\mathbb{R}^3_\lambda$ used in the physics literature. Note that the RHS of \eqref{def-com-rel2} is the Casimir operator for $su(2)$.\\

\subsection{\texorpdfstring{Noncommutative differential geometry set-up.}{Noncommutative differential set-up.}} \label{subsection22}

At the classical level, the construction of noncommutative gauge models can be done once a noncommutative differential calculus has been chosen. A particular version of the derivation-based differential calculus, a natural noncommutative extension of the usual de Rham complex, will be considered in the sequel. This has been introduced a long ago in \cite{mdv88-99}, inspired partly from the Koszul algebraic formulation of standard differential geometry \cite{koszul}. For mathematical developments and applications to noncommutative field theories see \cite{cgmw-20} and references therein. Informally, the key of this noncommutative differential calculus is to interpret the derivations of the algebra as the noncommutative analogs of the vector fields. Notice that the derivation based differential calculus does not exploit the natural Hopf algebra structure present on $\mathbb{R}^3_\lambda$. A possible choice would be to start from the bicovariant differential calculus \cite{woro}, which will not be considered here.\\

Let $\mathcal{G}$ be the Lie algebra of real inner derivations of $\mathbb{R}^3_\lambda$ defined as in \cite{gervitwal-13} by
\begin{equation}
\mathcal{G} := \{D_\alpha:= i [\theta_\alpha, \cdot]\} \ ,\ \ \theta_\alpha := \frac{x_\alpha}{\lambda^2} \ , \ \ \forall \alpha = 1,2,3 \ . \label{inv-form-conn}
\end{equation}
Thus, one has
\begin{equation}
[D_\alpha,D_\beta] = -\frac{1}{\lambda} \epsilon_{\alpha\beta\gamma} D_\gamma \ . \ \ \forall \alpha,\beta,\gamma = 1,2,3 \ . \label{Der}
\end{equation}
The resulting $\mathbb{N}$-graded differential algebra is $(\Omega_\mathcal{G}^\bullet = \oplus_{n\in\mathbb{N}} \Omega^n_\mathcal{G},\ d,\ \times)$, where $\Omega^n_\mathcal{G}$ is the space of $n-(\mathcal{Z}(\mathbb{R}^3_\lambda))$-linear) antisymmetric maps $\omega:\mathcal{G}^n\to\mathbb{R}^3_\lambda$, $\Omega^0_\mathcal{G}=\mathbb{R}^3_\lambda$ and $d:\Omega^n_\mathcal{G}\to\Omega^{n+1}_\mathcal{G}$ is the nilpotent differential defined 
for any $\omega\in\Omega^p_\mathcal{G}$ and $\rho\in\Omega^q_\mathcal{G}$ by
\begin{eqnarray}
d\omega(X_1,...,X_{p+1})&=&\sum_{k=1}^{p+1}(-1)^{k+1}X_k\omega(X_1,...,\vee_k,...,X_{p+1})\nonumber\\
&+&\sum_{1\le k<l\le p+1}(-1)^{k+l}\omega([X_k,X_l],...,\vee_k,...,\vee_l,...,X_{p+1})\label{differential},
\end{eqnarray}
in which the symbol $\vee_k$ means "element of rank $k$ omitted" and product $\times$ on $\Omega_\mathcal{G}^\bullet$ defined for any $\omega\in\Omega^p_\mathcal{G}$ and $\rho\in\Omega^q_\mathcal{G}$ by
\begin{equation}
\omega\times\rho(X_1,...,X_{p+q})=\frac{1}{p!q!}\sum_{\sigma\in\mathfrak{S}_{p+q}}\vert\sigma\vert\omega(X_{\sigma(1),...,X_{\sigma(p)}})
\rho(X_{\sigma(p+1),...,X_{\sigma(p+q)}})\label{innerformproduct},
\end{equation}
in which $X_i\in{\cal{G}}$'s, $\vert\sigma\vert$ is the signature of the permutation $\sigma\in\mathfrak{S}_{p+q}$.\\

Different notions of noncommutative connection have been introduced. Here, the notion of (hermitean) connection on a right-module over the algebra will be used \cite{mdv88-99} which is the one mostly used in the physics literature on the noncommutative field theories. It can be viewed informally as a noncommutative extension of the notion of connection on a module introduced by Koszul \cite{koszul} in the framework of standard differential geometry. Note that it would be interesting to carry out an analysis similar to the one presented below starting with the notion of connection over a bimodule \cite{michor-dbv} which should be more suited for a bicovariant differential calculus \cite{woro}.\\

Let $\mathbb{M}$ be a hermitean (right-)module over the algebra with hermitean structure{\footnote{A hermitean structure is defined as a sesquililear form $h:\mathbb{M}\times\mathbb{M}\to\mathbb{A}$ (here $\mathbb{A}=\mathbb{R}^3_\lambda$) with $h(ma,nb)=a^\dag h(m,n)b$, $h(m,m)\in\mathbb{A}_+$, $h(m,m)=0\Rightarrow m=0$, for any $a,b\in\mathbb{A}$ and any $m,n\in\mathbb{M}$.}} denoted by $h$. A hermitean connection on $\mathbb{M}$ can be defined as a linear map :
\begin{eqnarray}
\nabla&:&\mathbb{M}\to\mathbb{M}\otimes\Omega^1_{\mathcal{G}}\nonumber\\
\nabla(ma)&=&\nabla(m)a+m\otimes da,\nonumber\\ 
dh(m,n)&=&h(\nabla(m),n)+h(m,\nabla(n))\label{def-connection}
\end{eqnarray}
for any $m, n\in\mathbb{M}$ and any $a$ in the algebra. The group of gauge transformations $\mbox{Aut}(\mathbb{M},h)$ defined as the group of the automorphisms of $\mathbb{M}$ preserving $h$, i.e $h(\phi(m),\phi(n))=h(m,n)$, acts on the real affine space of hermitean connections as
\begin{equation}
\phi\triangleright\nabla:=\nabla^\phi=\phi^{-1}\circ\nabla\circ \phi,\label{gene-gauge}
\end{equation}
for any $\phi\in\mbox{Aut}(\mathbb{M},h)$. The curvature can then be defined as the morphism of module given by 
\begin{equation}
\nabla^2:\mathbb{M}\to\mathbb{M}\otimes \Omega^2_{\mathcal{G}}.\label{curvature}.
\end{equation}

Assuming now $\mathbb{M}=\mathbb{R}^3_\lambda$, $h(m_1,m_2)=m_1^\dag m_2$, one easily find that the corresponding hermitean connection is characterized by the 1-form $A:=\nabla(\bbone)\in\Omega^1_{\mathcal{G}}$ with related 2-form curvature $F=dA+A^2$, where in obvious notations
\begin{equation}
\nabla_{D_\mu}(a) := \nabla_\mu(a) = D_\mu a + A_\mu a,\ \ 
A_\mu :=\nabla_\mu(\bbone) \label{connection}
\end{equation}
with $A_\mu^\dag = - A_\mu$ for any $a\in\mathbb{R}^3_\lambda$ and 
\begin{equation}
F(D_\mu,D_\nu) := F_{\mu\nu} = [\nabla_\mu,\nabla_\nu] - \nabla_{[D_\mu,D_\nu]} = D_\mu A_\nu - D_\nu A_\mu + [A_\mu,A_\nu] + \frac{1}{\lambda} \epsilon_{\mu\nu\gamma} A_\gamma, \label{curv1}
\end{equation}
for $\mu = 1,2,3$. \\
The gauge transformations are given by
\begin{equation}
A_\mu^g = g^\dag A_\mu \ g + g^\dag D_\mu \ g,\ \ F^g_{\mu\nu} = g^\dag F_{\mu\nu} \ g \ .
\end{equation}
where $g:=\phi(\bbone)$, $\phi\in\mbox{Aut}(\mathbb{M},h)$, $g^\dag g=gg^\dag=\bbone$ so that the gauge group is the group of the unitary elements of the module $\mathcal{U}(\mathbb{R}^3_\lambda)$.\\

The space $\Omega^1_{\mathcal{G}}$ involves a distinguished element defined by
\begin{equation}
\Theta\in\Omega^1_{\mathcal{G}},\ \ \Theta(D_\mu):=\Theta_\mu=-i\theta_\mu\label{grandtheta}
\end{equation}
where $\theta_\mu$ has been given in \eqref{inv-form-conn}. By using \eqref{differential} and \eqref{innerformproduct}, one easily computes $d\Theta(D_\mu,D_\nu)=-\frac{1}{\lambda}\varepsilon_{\mu\nu\rho}\Theta_\rho$ and $\Theta\times\Theta(D_\mu,D_\nu)=\frac{1}{\lambda}\varepsilon_{\mu\nu\rho}\Theta_\rho$ so that
\begin{equation}
F^{inv}:=d\Theta+\Theta\Theta=0\label{flat-curv}.
\end{equation}
Moreover, let 
\begin{eqnarray}
i_X&:&\Omega^p_{\mathcal{G}}\to\Omega^{p-1}_{\mathcal{G}},\ (i_X\omega)(X_1,...,X_{p_1})=\omega(X,X_1,...,X_{p-1})\label{interior}\\
L_X&:&\Omega^p_{\mathcal{G}}\to\Omega^{p}_{\mathcal{G}},\ L_X=i_Xd+di_X\label{liederiv}, 
\end{eqnarray}
for any $X,\ X_k \in\mathcal{G},\ (k=1,2,...,p-1)$, denote in standard notations the Cartan operations for the action of the Lie algebra of derivations $\mathcal{G}$ on the graded differential algebra $\Omega^\bullet_{\mathcal{G}}$. $i_X$ and $L_X$ act as derivations respectively with $-1$ and $0$ degree. By noticing that one can write $d=-[\Theta,]$ on $\Omega^0_{\mathcal{G}}$, reflecting the fact that the derivations in $\mathcal{G}$ are inner, and making use of standard properties of graded commutators, one infers
\begin{equation}
L_X\Theta=i_Xd\Theta+d(i_X\Theta)=i_X(d\Theta+\Theta\Theta)=i_XF^{inv}
=0\label{invar},
\end{equation}
owing to \eqref{flat-curv} which therefore indicates that $\Theta$ is an invariant form in the language of Cartan operations but not horizontal (since one has $i_X\Theta\ne0$). Recall that Cartan operations appears as building ingredients in the formulation of invariant and basic cohomologies, which are essential to deal with topological field theories (of cohomological types) \cite{stor-wal} as well as in algebraic formulation of BRST symmetry and related s-cohomology modulo $d$ in connection with the algebraic classification of (chiral) anomalies \cite{RS}.\\
The invariant 1-form $\Theta$ defines actually the form-connection for the canonical gauge-invariant connection that occurs in the present set-up. From 
\eqref{def-connection}, it can be readily realized ($\nabla^{inv}(\bbone)=\Theta$) that this latter is given by
\begin{equation}
\nabla^{inv}(a):=da+\Theta a=a\Theta, \forall a\in\mathbb{R}^3_\lambda\label{inv-form-connection}.
\end{equation}
Gauge invariance follows immediately from \eqref{gene-gauge}. The corresponding curvature is given by \eqref{flat-curv} computed just above. Hence $\Theta$ defines a flat connection.\\
A natural gauge covariant tensor 1-form is then defined from
\begin{equation}
(\nabla-\nabla^{inv})(a)= (A-\Theta)(a):=\mathcal{A}(a)\label{tens-form-abst}
\end{equation}
for any $a\in\mathbb{R}^3_\lambda$ which satisfies for any $g\in\mathcal{U}(\mathbb{R}^3_\lambda)$
\begin{equation}
\mathcal{A}^g=g^\dag\mathcal{A}g\label{covar-tens-form}.
\end{equation}
This tensor form is sometimes related in the physics literature to the "covariant coordinates" which is apparent when expressing the "components" of the forms that we give below for further convenience. Namely,
\begin{equation}
\mathcal{A}_\mu = \nabla_\mu - \nabla^{inv}_\mu = A_\mu + i \theta_\mu,\ \forall i=1,2,3 \ , \label{tens-form}
\end{equation}
with $\mathcal{A}_\alpha^\dag = - \mathcal{A}_\alpha$, $\alpha=1,2,3$ ($A_\alpha^\dag=-A_\alpha$). 
In the same way
\begin{equation} 
\nabla^{inv}_\mu(a) = D_\mu a - i \theta_\mu a = - i a \theta_\mu,\ \forall a \in \mathbb{R}^3_\lambda\label{invar-connect}.
\end{equation}
One can check
\begin{equation}
\Theta_\mu^g=\Theta_\mu\label{invariance-theta},
\end{equation}
while the curvature takes the form
\begin{equation} 
F_{\mu\nu} = [\mathcal{A}_\mu,\mathcal{A}_\nu] + \frac{1}{\lambda} \epsilon_{\mu\nu\gamma} \mathcal{A}_\gamma \ . \label{curv2}.
\end{equation}
with 
\begin{equation}
F_{\mu\nu}^g =g^\dag F_{\mu\nu} g,
\end{equation}
for any $g\in\mathcal{U}(\mathbb{R}^3_\lambda)$.\\

\section{\texorpdfstring{Exact formulas in noncommutative gauge models on $\mathbb{R}^3_\lambda$.}{Exact formulas in noncommutative gauge models on $\mathbb{R}^3_\lambda$.}}\label{section3}

\subsection{\texorpdfstring{Noncommutative gauge (matrix) models.}{Noncommutative gauge (matrix) models.}} \label{section31}

From the above, it follows that any functional of the form $\tr(P(\mathcal{A}))$ where $P$ is a polynomial will give rise to a gauge invariant object. This simplifies the construction of candidates for classical gauge theory models and permits one to express/represent such a gauge theory model as a "matrix model" defined by a functional action $S(\mathcal{A})$ with $\mathcal{A}$ as field variable. For technical reasons, the study of such a matrix model representation is sometimes easier than its partner with $A_\mu$ as field variables. This observation has been widely exploited e.g in the context of theories on Moyal spaces $\mathbb{R}^4_\theta$ leading to interesting semi-classical investigations \cite{matrix1}-\cite{matrix5}. Whenever $S(\mathcal{A})$ supports $\mathcal{A}^0=0$ as some vacuum configuration, one may interpret $S(\mathcal{A})$ either as a model describing the dynamics of the fluctuations of $\mathcal{A}_\mu$ around $0$ or alternatively, in view of \eqref{tens-form-abst}, \eqref{tens-form}, as a model describing the fluctuations of $A_\mu$, the "gauge potential", around the "gauge-invariant connection". \\
As far as gauge theory models on $\mathbb{R}^3_\lambda$ as well as on Moyal spaces are concerned, it appears that a wide class of models has vacuum instabilities whenever the vacuum does not correspond to this gauge-invariant potential (see e.g \cite{MVW13}, \cite{gervitwal-13}). At the present time, I do not have any explanation (if any) of this behavior.\\

We close this subsection by noticing that the gauge invariant object $\Theta_\mu\Theta_\mu$ verifies
\begin{equation}
\Theta_\mu\Theta^\mu\in\mathcal{Z}(\mathbb{R}^3_\lambda)\label{harm-center}
\end{equation} 
which can be easily verified by using \eqref{grandtheta} together with \eqref{inv-form-conn} and \eqref{def-com-rel1}, \eqref{def-com-rel2} and recalling that $\mathcal{Z}(\mathbb{R}^3_\lambda)$ is generated by $x_0$. From this, it follows that for any polynomial $P(\mathcal{A})$, one has
\begin{equation}
\tr(P(\mathcal{A})\Theta_\mu\Theta^\mu)^g=\tr(P(\mathcal{A})\Theta_\mu\Theta^\mu)
\label{harmonic-polyn}
\end{equation}
i.e, it is gauge invariant, which can be realized by using gauge invariance of $\Theta_\mu$ together with \eqref{covar-tens-form}, \eqref{harm-center} and cyclicity of the trace.\\
We set from now on
\begin{equation}
\mathcal{A}_\mu=i\Phi_\mu.
\end{equation}
From \eqref{harmonic-polyn} and in view of \eqref{inv-form-conn}, one concludes that gauge invariant harmonic terms 
\begin{equation}
\sim\tr(x^2\Phi_\mu\Phi^\mu) \label{trap}
\end{equation}
are allowed in any gauge-invariant classical action on $\mathbb{R}^3_\lambda$. Note that harmonic terms pertain to the liturgy of noncommutative field theories. Such a term has been initially used as an essential ingredient in the construction of a all order perturbatively renormalisable scalar field theory with quartic interaction on the Moyal space $\mathbb{R}^4_\theta$ \cite{Grosse:2003aj-pc}. Its effect is, roughly speaking, to increase sufficiently the decay behavior of the propagator so that it can actually neutralize the so called UV/IR mixing that occurs on $\mathbb{R}^4_\theta$. \\
As far as gauge theories are concerned, a harmonic term $\sim\tr(x^2\Phi_\mu\Phi^\mu)$ would break gauge invariance in the $\mathbb{R}^4_\theta$ case because the counter part of \eqref{harm-center} does not hold true, namely $\Theta_\mu\Theta^\mu\notin\mathcal{Z}(\mathbb{R}^4_\theta)$ ($\Theta$ being here the gauge-invariant connection for $\mathbb{R}^4_\theta$, see e.g \cite{cgmw-20}).\\

Looking for positive gauge invariant actions (at most quartic in the field $\Phi_\mu$) that support $\Phi_\mu=0$ as classical vacuum configuration, the analysis in \cite{GJW2015} gave rise to the following family of classical actions
\begin{equation}
S_{cl}=\frac{1}{g^2}\tr\big([\Phi_\mu,\Phi_\nu]^2+\Omega\{\Phi_\mu,\Phi_\nu\}^2+(M+\mu x^2)\Phi_\mu\Phi^\mu \big)\label{classic action},
\end{equation}
with
\begin{equation}
\Omega\ge0,\ \mu>0,\ M>0
\end{equation}
where we set $x^2=x_\mu x^\mu$ with mass dimensions $[\Omega]=0$, $[\mu]=4$, $[M]=2$, $[g^2]=1$ so that the action is dimensionless, assuming that the relevant dimension here is the "engineering dimension" of $\mathbb{R}^3_\lambda$ equal to $3$. In \eqref{classic action}, $\{a,b\}:=ab+ba$. Notice that the action \eqref{classic action} is similar to the action 
for a matrix model.\par 

As shown in \cite{GJW2015}, the gauge-fixing can be conveniently performed in the gauge $\Phi_3=\theta_3$ while gauge-invariance of \eqref{classic action} is traded for invariance under a BRST symmetry \cite{BRS} whose structure equations defining the nilpotent Slavnov operation $s$ are \cite{GJW2015}
\begin{eqnarray}
s\Phi_\alpha &=& i [C,\Phi_\alpha],\ sC=iCC\label{brs1}\\
s{\bar{C}} &=& b,\ sb = 0 \label{contractible-brs}
\end{eqnarray}
where $C$ is the ghost field with ghost number $+1$ and ${\bar{C}}$ and $b$ are respectively the antighost and the St\"uckelberg field (with respective ghost number $-1$ and $0$). The Slavnov operation $s$ acts as an antiderivation with respect to the grading given by (the sum of) the ghost number (and degree of forms), modulo 2. \\
Then, adding to \eqref{classic action} the $s$-exact gauge-fixing action
\begin{equation}
S_{\phi\pi}=s\tr\big({\bar{C}}(\Phi_3-\theta_3) \big)=\tr\big(b(\Phi_3-\theta_3)-i{\bar{C}}[C,\Phi_3]\big)\label{gauge-fix}, 
\end{equation}
and integrating over the Stueckelberg field $b$ which enforces the gauge condition $\Phi_3=\theta_3$, it can be realized that the ghost part decouples. Finally, defining the complex fields
\begin{equation}
\Phi:=\frac{1}{2}(\Phi_1+i\Phi_2),\ \Phi^\dag:=\frac{1}{2}(\Phi_1-i\Phi_2),
\end{equation}
one obtains the following gauge-fixed action
\begin{equation}
S^f_\Omega = \frac{2}{g^2} \tr( \Phi \mathcal{Q} \Phi^\dag + \Phi^\dag \mathcal{Q}\Phi) + \frac{16}{g^2} \tr( (\Omega+1) \Phi\Phi^\dag\Phi\Phi^\dag + (3\Omega-1) \Phi\Phi\Phi^\dag\Phi^\dag),
\label{quasilsz}
\end{equation}
where the kinetic operator $\mathcal{Q}$ is an element of $\mathcal{L}(\mathcal{H})$, the space of linear operators acting on the Hilbert space 
\begin{equation}
\mathcal{H}=\mbox{span}\{v^j_{mn},\-j\le m,n\le j,\  j\in\frac{\mathbb{N}}{2}\}
\end{equation}
and is given by
\begin{equation}
\mathcal{Q}=M\bbone+\mu L(x^2)+8\Omega L(\theta_3^2)+i4(\Omega-1)L(\theta_3)D_3\label{kinetic-operator}
\end{equation}
in which $L(.)$ denotes the left multiplication. $\mathcal{Q}$ is self-adjoint as it can be seen by noticing that the first 3 terms in \eqref{kinetic-operator} are expressible as a sum a orthogonal projectors while self-adjointness of the last term stems from the self-adjointness of $L(\theta_3)$ and $D_3$ together with $[D_3,L(\theta_3)]=0$. \\
It is convenient to rewrite any $j$-component of the kinetic term as
\begin{equation}
\frac{2}{g^2}\mbox{tr}_j(\Phi\mathcal{Q}\phi^\dag+\Phi^\dag\mathcal{Q}\Phi):=\frac{1}{g^2}\sum_{\mu=1}^2\sum_{m,n,k,l}(\Phi^\mu)^j_{mn}(\Phi^\mu)^j_{kl}(\mathcal{Q})^j_{mn;kl}\label{qj-def}
\end{equation}
with
\begin{equation}
(\mathcal{Q})^j_{mn;kl}=\delta_{ml}\delta_{nk}\big(M+\mu\lambda^2j(j+1)+\frac{2\Omega}{\lambda^2}(k+l)^2
+\frac{2}{\lambda^2}(k-l)^2\big)\label{kin-matrix}.
\end{equation}
Then, the spectrum of $\mathcal{Q}\in\mathcal{L}(\mathcal{H})$ is given by
\begin{equation}
\mbox{spec}(\mathcal{Q})=\big\{M+\lambda^2\mu j(j+1)+
\frac{2\Omega}{\lambda^2}(k+l)^2+\frac{2}{\lambda^2}(k-l)^2,\-j\le k,l\le j,\ j\in\frac{\mathbb{N}}{2} \big\}\label{spectrumQ}
\end{equation}
with finite degeneracy for each of the eigenvalues which decays to $0$ as $j\to\infty$. Hence, the resolvent operator of $\mathcal{Q}$, $R_{\mathcal{Q}}(z)=(\mathcal{Q}-z\bbone)^{-1}$, for any $z\notin\mbox{spec}(\mathcal{Q})$ is compact: $R_{\mathcal{G}}(z)\in\mathcal{K}(\mathcal{H})$. Finally, the spectrum \eqref{spectrumQ} is positive which implies that $\mathcal{Q}$ is a positive self-adjoint operator.\\

The main result of \cite{GJW2015} holds true for the action \eqref{quasilsz}. Namely, one has the following property:
\begin{theorem} [\cite{GJW2015}]
The amplitudes of the ribbon diagrams of any arbitrary order for the functional action \eqref{quasilsz} for $M>0$, $\mu>0$, $\Omega>0$ are finite.
\end{theorem}
The somewhat lengthy proof given in \cite{GJW2015} can be achieved thanks in particular to a power counting for the ribbon diagrams stemming from the perturbative expansion. At this point, some comments are in order. First, the UV (and IR) finiteness of the gauge theory model \eqref{quasilsz} actually stems from the combination of:\\
i) the existence of an upper bound for the propagator $\mathcal{Q}^{-1}$ which by the way corresponds to the propagator of another all order finite gauge invariant model that I will not discuss here (see \cite{GJW2015}),\\
ii) the salient role played by $j\in\frac{\mathbb{N}}{2}$ which acts as a natural (UV) cut-off,\\
iii) a sufficient rapid decay of the propagator at large $j$ (corresponding to the UV region) insured by the presence of the gauge-invariant harmonic term discussed above.\\
Note that the action \eqref{quasilsz}, which can be viewed as describing the fluctuations of the covariant coordinates around the vacuum $\Phi^0=0$ can be alternatively interpreted as describing the dynamics of the fluctuations of the gauge potential $A_\mu$ around the "gauge-invariant connection", says $A_\mu^0=-\theta_\mu$, since the covariant coordinates are defined as the difference of 2 connections, as discussed above. \\
Next, it can be realized that the origin of property ii) given above stems from the Peter-Weyl decomposition of the algebra. Hence, I expect that a similar feature (namely the occurrence of natural UV cut-offs) should hold true in generalization of the present construction of other compact group (e.g $SU(n)$). \par

\subsection{\texorpdfstring{Partition functions as ratios of determinants.}{Partition functions as ratios of determinants.}}\label{subsection32}

I assume from now on $\Omega=\frac{1}{3}$. Accordingly, the last interaction term vanishes so that the quartic interaction term depends only on $\Phi\Phi^\dag$. The action \eqref{quasilsz} reduces to
\begin{equation}
S^f_{1/3} = \frac{2}{g^2} \tr( \Phi {Q} \Phi^\dag + \Phi^\dag {Q}\Phi) + \frac{64}{3g^2} \tr(\Phi\Phi^\dag\Phi\Phi^\dag)\label{action-int},
\end{equation}
where the positive self-adjoint operator $Q$ is
\begin{equation}
Q=M\bbone+\mu L(x^2)+\frac{8}{3} L(\theta_3^2)-i\frac{8}{3}L(\theta_3)D_3\label{kinetic-Q}.
\end{equation}
One observes that the action {\it{formally}} shares some common points with the action describing an exactly solvable model investigated in \cite{LSZ}.  It turns out that the partition function for $S^f_{1/3}$ \eqref{quasilsz} can be related to $\tau$-functions of integrable hierarchies. \\ 

Indeed, thanks to the Peter-Weyl decomposition of $\mathbb{R}^3_\lambda$ \eqref{convolution-alg}, the partition function can be expressed as a product of factors labeled by $j\in\frac{\mathbb{N}}{2}$, each one related to a ratio of determinants. Note that each of these factors can be interpreted as the partition function for the reduction of the gauge-fixed theory \eqref{quasilsz} on the matrix algebra $\mathbb{M}_{2j+1}(\mathbb{C})$, i.e a fuzzy sphere of radius $j$. A standard computation using $Q$ \eqref{kinetic-Q} gives rise to the following expression for the partition function 
\begin{equation}
Z(Q) = \prod_{j\in\frac{\mathbb{N}}{2}} Z_j(Q)\label{zq},
\end{equation}
where
\begin{eqnarray}
Z_j(Q) &=&\int{\mathcal{D}} \Phi^j \mathcal{D}\Phi^{\dag j} \ exp(-S_j(\Phi,\Phi^\dag,Q)),\nonumber\\
S_j(\Phi,\Phi^\dag,Q)&=&\frac{w(j)}{g^2} ( 2 \mbox{ tr}_j(\Phi^j Q^j\Phi^{\dag j}+\Phi^{\dag j} Q^j\Phi^j)+\frac{64}{3}\mbox{ tr}_j(\Phi^j\Phi^{\dag j}\Phi^j\Phi^{\dag j}) ) \label{zqj}
\end{eqnarray}
with
\begin{equation}
\mathcal{D} \Phi^j \ \mathcal{D} \Phi^{\dag j} := \prod_{-j\le m,n\le j} \mathcal{D} \Phi^j_{mn} \mathcal{D} \Phi^{\dag j}_{mn} \ , \label{measure}
\end{equation}
and we set
\begin{equation}
w(j)=8\pi\lambda^3(2j+1).
\end{equation}
The matrix $Q^j\in\mathbb{M}_{2j+1}(\mathbb{C})$ can be obtained from \eqref{kinetic-Q} and \eqref{kin-matrix} defining the operator $Q\in\mathcal{L}(\mathcal{H})$ for $\Omega=\frac{1}{3}$, namely
\begin{equation}
({Q})^j_{mn;kl}=\delta_{ml}\delta_{nk}\big(M+\mu\lambda^2j(j+1)+\frac{2}{3\lambda^2}(k+l)^2
+\frac{2}{\lambda^2}(k-l)^2\big)\label{kin-matrix-Q}.
\end{equation}
$Z_j(Q)$ for any $j\in\frac{\mathbb{N}}{2}$ can be interpreted as the partition function for the gauge model truncated to a "fuzzy sphere" $\mathbb{M}_{2j+1}(\mathbb{C})$. It turns out that the functional integration in \eqref{zqj} can be entirely performed. As a result, $Z_j(Q)$ is expressible as a ratio of determinants, up to an unessential prefactor, as I now show.\\

Define a change of integration variable by making use of a singular value decomposition of $\Phi^j$. Namely, one has
\begin{equation}
\Phi^j=U^\dag R^jV, 
\end{equation}
where $U$ and $V$ are unitary matrices in $\mathbb{M}_{2j+1}(\mathbb{C})$ and $R^j\in\mathbb{M}_{2j+1}(\mathbb{C})$ is a diagonal positive matrix. Set 
\begin{equation}
R^j:=\text{diag}(\rho^j_{m}),\ \rho^j_m\ge0
\end{equation}
and
\begin{equation}
t^j_m:=(\rho^{j}_m)^2,
\end{equation}
for any $-j\le m\le j$. \\
Let $d\theta(X)$ denotes the invariant Haar measure of the unitary group $U(2j+1)$ for any $X\in U(2j+1)$. Using the Jacobian for the above change of variables defined by
\begin{equation}
\mathcal{D} \Phi^j\mathcal{D} \Phi^{\dag j}=\Delta^2(R^{j2})d\theta(U)d\theta(V) \prod_{k=-j}^j dt^j_k\label{jacobianU}
\end{equation}
where $\Delta(R^{j2})$ denotes the Vandermonde determinant related to the matrix $R^{j2}$ given by
\begin{equation}
\Delta(R^{j2}) = \prod_{-j\le k<l\le j} (t^{j}_l-t^{j}_k), \label{vanderrho2}
\end{equation}
the partition function \eqref{zqj} can be cast into the form
\begin{eqnarray}
Z_j(Q)&=&\int_0^{+\infty}\prod_{k=-j}^j dt^j_k\Delta^2\left(R^{j2}\right)\int_{U(2j+1)} d\theta(U)d\theta(V)
e^{-S_j(Q;U,V,R)}\nonumber\\
S_j(Q;U,V,R)&=&\frac{w(j)}{g^2}(2\mbox{ tr}_j(VQ^jV^\dag R^{j2}+UQ^jU^\dag R^{j2})+\frac{64}{3}\mbox{ tr}_j(R^{j4}))\label{zj-interm}.
\end{eqnarray}
From \eqref{zj-interm}, one observes that one can decouple the field variables $U$ and $V$ (the "angular" part) from the positive diagonal ("radial") part, thanks to the expression for the quartic potential at $\Omega=\frac{1}{3}$ (see \eqref{quasilsz}). \\
Indeed, the integration over $U$ and $V$ can be performed by using the Harish-Chandra/Itzykson-Zuber measure formula of the random matrix theory. Recall that for any hermitean matrices $M,N\in\mathbb{M}_n(\mathbb{C})$ with eigenvalues of $M$ ordered as $\lambda^M_1\le\lambda^M_2\le...\le\lambda^M_n$ (and similar ordering for $N$) and any unitary matrix $U\in\mathbb{M}_n(\mathbb{C})$, the following formula holds true:
\begin{equation}
\int_{U(n)} d\theta(U) \ e^{z\mbox{ tr}(MUNU^\dag)} = \frac{1}{\Delta(M)\Delta(N)} \ \prod_{k=1}^{n-1}k! \ z^{\frac{n(1-n)}{2}} \ \det_{1\le k,l\le n}(e^{z\lambda^M_k\lambda^N_l}), \label{HC}
\end{equation}
for any $z\in\mathbb{C} \backslash \{0\}$, $d\theta(U)$ is the Haar measure on $U(n)$ and $\Delta(M)$, $\Delta(N)$ are the Vandermonde determinants for $M$ and $N$ as defined above.

Using \eqref{HC} in \eqref{zj-interm} yields

\begin{eqnarray}
Z_j(Q) &=& \frac{N^j(g^2)}{\Delta^2(Q^j)} \int_0^\infty \prod_{k=-j}^j dt^j_k\left( \det_{-j\le p,l\le j}\left(e^{-2\frac{w(j)}{g^2}t^j_p\omega^j_{l}}\right)\right)^2 \ e^{-\frac{64w(j)}{3g^2} \underset{-j<m<j}{\sum} \ t^{j2}_m} \ , \label{zj-partial} \\
&=& \frac{N^j(g^2)}{\Delta^2(Q^j)} \int_0^\infty \prod_{k=-j}^j dt^j_k \left( \sum_{\sigma\in\mathfrak{S}_{2j+1}} \left|\sigma\right| \prod_{k=-j}^j \ e^{-\frac{2w(j)}{g^2}t^j_k\omega^j_{\sigma(k)}} \right)^2 \ e^{-\frac{64w(j)}{3g^2} \underset{m}{\sum} t^{j2}_m} \ , \nonumber \label{zj-partialbis} \\
&&
\end{eqnarray}
where
\begin{equation}
N^j(g^2) = \left(\prod_{k=1}^{2j}k!\right)^2 \left(\frac{2w(j)}{g^2}\right)^{-2j(2j+1)} \ , \label{normal-zj}
\end{equation}
and $\omega^j_k$ are the eigenvalues of the real symmetric matrix $Q^j\in\mathbb{M}_{2j+1}(\mathbb{C})$ which is related to \eqref{kin-matrix-Q}. In the second expression for $Z_j(Q)$ \eqref{zj-partialbis}, $\vert\sigma\vert$ denotes the signature of the permutation $\sigma$ in $\mathfrak{S}_{2j+1}$.\\

The integration over the $t^j_k$'s can now be performed. We expand the square of the sum in \eqref{zj-partialbis} to obtain
\begin{equation}
Z_j(Q) = \frac{N^j(g^2)}{\Delta^2(Q^j)} \ \sum_{\sigma_1,\sigma_2\in\mathfrak{S}_{2j+1}} |\sigma_1| \ | \sigma_2 | \prod_{k=-j}^j \int_0^\infty dt^j_k \ ( e^{-\frac{64w(j)}{3g^2} \underset{m}{\sum} \ t^{j2}_m} \ e^{-2\frac{w(j)}{g^2} t^j_k \omega^j_{\sigma_1\sigma_2(k)} } ) \ , \label{calcul1} 
\end{equation}
where we have defined
\begin{equation}
\omega^j_{\sigma_1\sigma_2(k)} := \omega^j_{\sigma_1(k)} + \omega^j_{\sigma_2(k)}. 
\end{equation}
We now combine \eqref{calcul1} with the relation
\begin{equation}
\int_0^\infty dxe^{-Ax^2-bx} = \sqrt{\frac{\pi}{2A}} \ \ \mbox{erfc}(\frac{b}{2\sqrt{A}}) \ e^{\frac{b^2}{4A}} \ , \quad \mbox{with } \ \Re(A)\ge0 \ , \ \ \Re(b)>0 \ , \label{erfc-integ}
\end{equation}
where ${\text{erfc}}$ is the complementary error function defined by
\begin{equation}
\text{erfc}(z) = \frac{2}{\sqrt{\pi}} \int_z^\infty dx \ e^{-x^2} \ , \quad \forall z \in \mathbb{R} \ ,
\end{equation}
to write $Z_j(Q)$ as
\begin{equation}
Z_j(Q) = \frac{N^j(g^2)}{\Delta^2(Q^j)} \sum_{\sigma_1,\sigma_2\in\mathfrak{S}_{2j+1}} \left|\sigma_1\right| \ \left|\sigma_2\right| \ \prod_{k=-j}^j f(\omega_{\sigma_1\sigma_2(k)}) \ , \label{calcul2}
\end{equation}
where
\begin{equation}
f(\omega_{\sigma_1\sigma_2(k)}) = \sqrt{\frac{\pi g^2}{128w(j)}} \ \ \text{erfc}\left(\sqrt{\frac{w(j)}{64g^2}} \ \ \omega^j_{\sigma_1\sigma_2(k)} \right) \ \ e^{\frac{w(j)}{64g^2} \ \omega^{j2}_{\sigma_1\sigma_2(k)}}.\label{functionspect}
\end{equation}
Finally, by using the properties of determinants, \eqref{calcul2} can be written as
\begin{equation}
Z_j(Q) = (N^j(g^2) \ (2j+1)!)\frac{\det_{-j\le m,n\le j} (f(\omega^j_m+\omega^j_n))}{\Delta^2(Q^j)} \ , \label{partitionfactor}
\end{equation}
for any $j\in\frac{\mathbb{N}}{2}$. Hence, all the functional integrals in $Z_j(Q)$ can be explicitly carried out so that any corresponding truncated gauge model on $\mathbb{M}_{2j+1}(\mathbb{C})$ can be viewed as an exactly solvable model.\\

The ratio of determinants appearing in the RHS of \eqref{partitionfactor} is somehow reminiscent of a $\tau$-function such as those occurring in integrable hierarchies. In fact,  \eqref{partitionfactor} could have been expected owing to the similarity between the present gauge model \eqref{action-int} and the LSZ model \cite{LSZ}. Recall that this latter belongs to a particular class of scalar field theories with quartic interaction built on the Moyal space $\mathbb{R}^4_\theta$. It has been shown to be exactly solvable, by exploiting a correspondence with the large $N$
limit of a complex 1-matrix model. \\
In the present situation however, one cannot take advantage of some large $N$ (i.e large $j$) limit to draw general conclusion on $Z_j(Q)$ at 
arbitrary $j$ (except for the case $j\to\infty$) and so on $Z(Q)=\prod_j Z_j(Q)$ \eqref{zq}. Recall that the other family of gauge models on $\mathbb{R}^3_\lambda$ investigated perturbatively in \cite{gervitwal-13}, when truncated to a single "fuzzy sphere" $\mathbb{M}_{2j+1}(\mathbb{C})$, is related to the Alekseev-Recknagel-Schomerus action \cite{ARS} which pertains to the area of string theory and describes the low energy action for brane dynamics on $\mathbb{S}^3$. It would be interesting to examine if some relation similar to \eqref{partitionfactor} shows up within some of these latter gauge model for some particular choice of parameters.\\

A relation between any truncated gauge model on $\mathbb{M}_{2j+1}(\mathbb{C})$  and integrable 2-D Toda lattice hierarchy can be conveniently exhibited by introducing in the partition function $Z_j(Q)$ a source term linearly coupled to the trace of the operator $\Phi^\dag\Phi$ (which may be viewed as a kind of analog of the condensate operator) i.e supplementing the argument of the exponential in \eqref{zqj} by 
$-\frac{w(j)}{g^2}\mbox{tr}_j(\Sigma^j\Phi^{\dag j} \Phi^j)$ where $\Sigma^j\in\mathbb{M}_{2j+1}(\mathbb{C})$ is the hermitean source of the "composite operator" (see \eqref{condensat-generating} of the appendix \ref{appendix}). Then, the corresponding partition function $Z_j(Q;\Sigma)$ can be expressed as
\begin{equation}
Z_j(Q;\Sigma) = \det_{-j\le m,n\le j} \big[\int \frac{dz_1}{i2\pi} \ \frac{dz_2}{i2\pi} \ z_1^{m-1} z_2^{n-1}f(z_1^{-1}+z_2^{-1}) e^{(\sum_{n=1}^\infty t_nz_1^n+ \bar{t}_nz_2^n)} \big], \label{tau-toda-2d}
\end{equation}
with
\begin{equation}
t_n = \frac{1}{n} \sum_{k=1}^{2j+1} (\omega^j_k)^n \ , \quad \bar{t}_n=\frac{1}{n}\sum_{k=1}^{2j+1}(\omega^j_k+\sigma^j_k)^n \label{time-variables} 
\end{equation}
in which $\sigma^j_k$, $-j\le k\le j$ are the eigenvalues of $\Sigma^j$ and $f$ is still given by \eqref{functionspect}. Equation \eqref{tau-toda-2d} corresponds to a $\tau$-function $\tau(t,\bar{t})$ for an integrable 2-D Toda lattice hierarchy. Setting $\Sigma^j=0$ in \eqref{tau-toda-2d} leads to the related expression for $Z_j(Q)$ which thus corresponds to a reduction of this hierarchy.\par

\section{Discussion and conclusion.}\label{section4}

$\mathbb{R}^3_\lambda$ as defined by \eqref{convolution-alg} supports a family of (matrix) gauge theory models described by \eqref{quasilsz} with stable vacuum which are perturbatively finite to all orders. The "mass term" $\Theta_\mu\Theta^\mu$ for the gauge-invariant connection $\Theta_\mu$ belongs to the center of the algebra insuring that gauge-invariant harmonic terms can be included in the functional action, thus implying that the gauge propagator decays as an inverse power of the natural UV cut-off $j$. The fact that $j$, the radius of the fuzzy sphere $\mathbb{M}_{2j+1}(\mathbb{C})$, plays the role of a UV cut-off comes from the Peter-Weyl decomposition of $\mathbb{R}^3_\lambda$ which enforces a factorization of the partition function as $Z(Q) = \prod_{j\in\frac{\mathbb{N}}{2}} Z_j(Q)$ \eqref{zq}, \eqref{zqj} where $Z_j(Q)$ can be viewed as the partition function for the gauge theory truncated on the fuzzy sphere $\mathbb{M}_{2j+1}(\mathbb{C})$. For a particular value of one parameter, namely $\Omega=\frac{1}{3}$, the quartic interaction term simplifies leading to \eqref{action-int} and each $Z_j(Q)$ can be exactly expressed as a ratio of determinants indicating that the corresponding truncated gauge theory is formally exactly solvable. A relation with (reduction of)  integrable 2-D Toda lattice hierarchy is also given. Hence, the gauge theory described by \eqref{action-int} is related to an infinite tower of solvable gauge theories on fuzzy spheres. A full characterization of the gauge theory \eqref{action-int} would need to carry out the resummation of $W(Q)=\sum_j\ln(Z_j(Q))$, which is not easy to achieve. This task has been undertaken.\\
The understanding of the quantum properties of NCFT and their gauge theoretic versions is still in its prime infancy, despite many advances achieved since the beginning of this century, obtained from the analysis of several representative prototypes mentioned or analyzed in this paper. These advances are mainly technical in nature, ranging from diagrammatic computational tools to adapted rules for "power counting". In their present formulation, (most of) these non local theories are rooted in an Euclidean set-up, stemming from the underlying noncommutative structures. A proper inclusion of some noncommutative analog of causality is needed in order to widen their possible relevance to physics and to understand what in NCFT supercede (at least) the concepts (and their interplays) of locality, microcausality and power counting ruling ordinary quantum field theories. These 3 notions were often present in numerous endless discussions I had with Raymond Stora so many years ago.

\vskip 1 true cm
{\bf{Acknowledgements:}} This work is dedicated to the memory of my colleague and friend Raymond Stora.  Discussions with M. Dubois-Violette, N. Franco, L. Landi and F. Latr\'emoli\`ere at various stages of this work are gratefully acknowledged.
\setcounter{section}{0}
\appendix
\section{\texorpdfstring{A link to integrable 2-D Toda lattice hierarchy.}{A link to integrable 2-D Toda lattice hierarchy.}}\label{appendix}
Write $\Phi^\dag\Phi=\sum_{j,m,n}(\Phi^\dag\Phi)^j_{mn}v^j_{mn}$ in obvious notations. Now, observe that the connected part of the expectation $\langle(\Phi^\dag\Phi) \rangle$ is determined by the quantities
\begin{equation}
\left\langle(\Phi^\dag\Phi)^k_{nm}\right\rangle = \frac{1}{Z(Q)} \ \frac{\delta}{\delta\Sigma^k_{nm}} \left(\prod_{j\in\frac{\mathbb{N}}{2}}Z_j(Q;\Sigma^j)\right)\Bigg|_{\Sigma=0 },\ k\in\frac{\mathbb{N}}{2} \ , \quad -k\le m,n\le k \ ,
\end{equation}
where
\begin{equation}
Z_j(Q;\Sigma) = \int{\mathcal{D}} \Phi^j \ \mathcal{D} \Phi^{\dag j} \ e^{-\frac{w(j)}{g^2} (2\mbox{tr}_j(\Phi^j Q^j\Phi^{\dag j}+\Phi^{\dag j} Q^j\Phi^j)+\frac{64}{3}\mbox{tr}_j(\Phi^j\Phi^{\dag j}\Phi^j\Phi^{\dag j})+\mbox{tr}_j(\Sigma^j\Phi^{\dag j}\Phi^j))}, \label{condensat-generating}
\end{equation}
and the source of the "composite operator" $\Sigma^j\in\mathbb{M}_{2j+1}(\mathbb{C})$ is hermitean. Hence, one can write $\Sigma^j=U\sigma^jU^\dag$ for some unitary matrix $U$ where $\sigma^j=\text{diag}(s^j_k)_{-j\le k\le j}$. From this follows that
\begin{equation}
\left\langle (\Phi^\dag\Phi)^j_{nm} \right\rangle = \frac{1}{Z_j(Q)} \ \frac{\delta}{\delta\Sigma^j_{mn}} Z_j(Q;\Sigma^j)\bigg|_{\Sigma^j=0} \ . \label{condensat-component}
\end{equation}
for any $j\in\frac{\mathbb{N}}{2}, \-j\le m,n\le j$. Moving to the sources $s^j_k$, it can be realized that the action of the functional derivative $\frac{\delta}{\delta s^j_k}$ generates the expectation $\langle (\Phi^\dag\Phi)^j_{qr}U_{rk}U^\dag_{kq}\rangle$ (no summation over $k$). Therefore%
\begin{equation}
\left\langle \mbox{tr}_j \left((\Phi^\dag\Phi)^j\right) \right\rangle = \sum_{k=-j}^j\frac{\delta}{\delta s^j_k}\ln(Z_j(Q;\Sigma^j)) \ ,
\end{equation}
where we used $U^\dag U=U^\dag=\bbone$. Now by performing a singular value decomposition of $\Phi^j$ in \eqref{condensat-generating} and integrating over the angular part 
using \eqref{HC}, we obtain
\begin{eqnarray}
Z_j(Q;\Sigma) &=& \frac{N^j(g^2)}{\Delta(Q^j)\Delta(Q^j+\Sigma^j)} \int_0^\infty\prod_{k=-j}^jdt^j_k \ \det_{-j\le p,l\le j}\left(e^{-2\frac{w(j)}{g^2}t^j_p\omega^j_{l}}\right) \nonumber \\
&\times& \det_{-j\le p,l\le j}\left(e^{-2\frac{w(j)}{g^2}t^j_p(\omega^j_{l}+\sigma^j_l)}\right) \ 
e^{-\frac{64w(j)}{3g^2} \underset{-j\leq m \leq j}{\sum} t^{j2}_m} \ , \label{decadix}
\end{eqnarray}
where $N^j(g^2)$ is still given by \eqref{normal-zj} and $\Delta(Q^j+\Sigma^j)$ is the Vandermonde determinant built from
\begin{equation}
\lambda^j_k=\omega^j_k+\sigma^j_k. 
\end{equation}
Expanding the determinants in the numerator of \eqref{decadix}, we obtain
\begin{eqnarray}
Z_j(Q;\Sigma)&=& \frac{N^j(g^2)}{\Delta(Q^j)\Delta(Q^j+\Sigma^j)} \sum_{\pi_1,\pi_2\in\mathfrak{S}_{2j+1}} \left|\pi_1\right| \ \left|\pi_2\right| \prod_{k=-j}^j f\left(\omega^j_{\pi_1(k)}+\Lambda^j_{\pi_2(k)}\right) \nonumber\\
&=& \frac{N^j(g^2) \ (2j+1)!}{\Delta(Q^j)\Delta(Q^j+\Sigma^j)} \ \det_{-j\le m,n\le j}\left(f\left(\omega^j_m+\Lambda^j_n\right)\right) \ , \label{generat-composit}
\end{eqnarray}
where $f(x)$ can be read off from \eqref{functionspect}.\par
It can be realized that the generating functional is given (up to the unessential overall factor $N^j(g^2)$ that we drop from no on)  ) to the $\tau$-function of an integrable 2-d lattice Toda hierarchy. Indeed, by using the the standard expression for the Vandermonde determinants in \eqref{generat-composit} as%
\begin{equation}
\Delta(x)=\det_{-j\le m,n\le j}\left(x_m^{n-1}\right) \ ,
\end{equation}
and reexpressing the ratio of determinants in \eqref{generat-composit} from a combination of complex integrals with the Cauchy-Binet identity given generically by
\begin{equation}
\exp\left(\sum_{n=1}^\infty t_nz^n\right) = \prod_{n=1}^{N}\frac{\lambda_n}{\lambda_n-z} \ , \quad t_n:=\frac{1}{n}\sum_{k=1}^{N}(\lambda_k)^n \ ,
\end{equation}
$Z_j(Q;\Sigma)$ can be easily cast into the form
\begin{equation}
Z_j(Q;\Sigma) = \det_{-j\le m,n\le j} \left(\int \frac{dz_1}{i2\pi} \ \frac{dz_2}{i2\pi} \ z_1^{m-1} z_2^{n-1} \ f\left(z_1^{-1}+z_2^{-1}\right) \ \exp\left(\sum_{n=1}^\infty t_nz_1^n+ \bar{t}_nz_2^n\right) \right) \ , \label{tau-toda}
\end{equation}
in which%
\begin{equation}
t_n = \frac{1}{n} \sum_{k=1}^{2j+1} (\omega^j_k)^n \ , \quad \bar{t}_n=\frac{1}{n}\sum_{k=1}^{2j+1}(\omega^j_k+\sigma^j_k)^n \ . \label{time-variable} 
\end{equation}

\vskip 1 true cm

\end{document}